\documentclass[a4paper,11pt]{article}
\pdfoutput=1
\usepackage[english]{babel}
\usepackage{jheppub}
\usepackage{subfigure}
\usepackage{xspace}
\usepackage{amsmath,amssymb,textcomp,enumerate,alltt,xspace,multirow}
\usepackage{slashed}
\usepackage{multicol}

\usepackage{graphicx}
\usepackage{relsize}
\usepackage{pstricks}
\usepackage{color}
\usepackage{xcolor}

\newcommand{\ri}{\mathrm{i}}

\newcommand{\calA}{\mathcal{A}}
\newcommand{\calC}{\mathcal{C}}

\newcommand{\ord}{\mathcal{O}}
\newcommand{\calO}{\ord}

\newcommand{\calW}{\mathcal{W}}

\newcommand{\ie}{i.e.~}

\newcommand{\fin}{\mathrm{fin}}

\newcommand{\epsfive}{\epsilon_5}
\newcommand{\eps}{\epsilon}
\newcommand{\asontwopi}{\frac{\as}{2\pi}}

\newcommand{\lb}{\left(}
\newcommand{\rb}{\right)}
\newcommand{\lsb}{\left[}
\newcommand{\rsb}{\right]}

\newcommand{\aew}{\alpha}
\newcommand{\as}{\alpha_\mathrm{s}}

\newcommand{\aspi}{\frac{\as}{2\pi}}
\newcommand{\nfvo}{n^{\gamma}_f}
\newcommand{\nfvt}{n^{\gamma\gamma}_f}

\def\refeq#1{\mbox{(\ref{#1})}}
\def\refeqs#1#2{\mbox{(\ref{#1})--(\ref{#2})}}
\def\reffi#1{\mbox{Fig.~\ref{#1}}}

\def\refta#1{\mbox{Table~\ref{#1}}}

\def\refse#1{\mbox{Section~\ref{#1}}}

\def\citere#1{\mbox{Ref.~\cite{#1}}}

\preprint{
\begin{flushright}
MSUHEP-21-005 \\
OUTP-21-04P   \\
\end{flushright}
}

\title{Two-loop leading colour QCD corrections to $q \bar{q} \to \gamma \gamma g$
and $q g \to \gamma \gamma q$}

\author[a]{Bakul Agarwal,}
\author[b]{Federico Buccioni,}
\author[a]{Andreas von Manteuffel,}
\author[b]{Lorenzo Tancredi}
\emailAdd{agarwalb@msu.edu}
\emailAdd{federico.buccioni@physics.ox.ac.uk}
\emailAdd{vmante@msu.edu}
\emailAdd{lorenzo.tancredi@physics.ox.ac.uk}

\affiliation[a]{
Department  of  Physics  and  Astronomy,  Michigan  State  University,\\
East  Lansing,  Michigan  48824,  USA
}

\affiliation[b]{
Rudolf Peierls Centre for Theoretical Physics, University of Oxford,\\
Clarendon Laboratory, Parks Road, Oxford OX1 3PU
}

\abstract{
We present the leading colour and light fermionic planar two-loop corrections for the production of two photons and a jet in the quark-antiquark and quark-gluon channels.
In particular, we compute the interference of the two-loop amplitudes with the corresponding tree level ones, summed over colours and polarisations.
Our calculation uses the latest advancements in the algorithms for
integration-by-parts reduction and 
multivariate partial fraction decomposition to produce compact and easy-to-use results.
We have implemented our results in an efficient C++ numerical code. 
We also provide their analytic expressions in Mathematica format.
}


\begin{document}
\maketitle

\flushbottom

\section{Introduction}
\label{se:intro}

The production of a pair of isolated highly-energetic photons in proton-proton collisions, $p p \to \gamma \gamma + X$,
represents an important class of processes for the physics programme at the Large Hadron Collider (LHC)
and at future hadron colliders.
Among many relevant aspects, pairs of prompt photons (diphotons) 
constitute an irreducible background to various Standard Model (SM) 
and Beyond Standard Model (BSM) processes, most prominently
Higgs boson production in its $H \to \gamma\gamma$ decay channel.
Indeed, with the increase of collision energy, 
the diphoton invariant mass distribution
can provide a powerful tool to search for heavy resonances
decaying to pairs of photons~\cite{Aaboud:2016tru,Sirunyan:2018wnk}, while
its transverse momentum distribution
offers a unique probe to investigate their properties. 
It is therefore crucial to provide an accurate theoretical description of 
the production of a pair of photons recoiling against hard Quantum Chromodynamics (QCD)
radiation, across a vast spectrum
of energies.

At the technical level,
the theoretical description of the production of a pair of photons with large transverse momentum 
is non trivial for several reasons.
An obvious one is that it requires computation of
$2 \to n$ scattering amplitudes with $n \geq 3$ for various partonic channels 
relevant at a specific perturbative order.
In particular, at leading order (LO) the process receives contribution from three partonic sub-channels, 
$q\bar{q} \to g \gamma \gamma$, $q g \to q \gamma \gamma $, and $\bar{q} g \to \bar{q} \gamma \gamma$, 
where the second and third channels can be obtained as crossings of the first.
The loop-induced gluon-fusion process $gg \to g \gamma \gamma$, instead, contributes formally only at next-to-next-to-leading-order (NNLO).

The mathematical complexity of loop corrections to the scattering amplitudes above is 
the main reason why
$pp \to \gamma \gamma + {\rm jet}$ is currently known only up to next-to-leading-order (NLO) QCD~\cite{DelDuca:2003uz,Gehrmann:2013aga}.
Calculation of cross section and differential distributions for 
$pp \to \gamma \gamma + {\rm jet}$ through NNLO QCD requires the computation of two-loop amplitudes
for $q\bar{q} \to g \gamma \gamma$ and its crossings, 
together with an efficient subtraction scheme to organise and cancel the infrared (IR) 
divergences between real and virtual contributions.
In recent years, a lot of progress has been achieved on both fronts. 
On the IR subtraction side, several schemes have been developed which are able to handle NNLO QCD
corrections to the production of a colour singlet plus one QCD jet~\cite{GehrmannDeRidder:2005cm,Daleo:2006xa,Czakon:2010td,Czakon:2014oma,Gaunt:2015pea,Boughezal:2015aha,Caola:2017dug}.
On the amplitude side, equally impressive results have been achieved. 
The first two-loop $2 \to 3$ scattering amplitudes for the production of three jets and three photons
in leading-colour QCD have been computed~\cite{Badger:2017jhb,Abreu:2017hqn,Abreu:2018zmy,Abreu:2018jgq,Chicherin:2018yne,Badger:2018enw,Badger:2019djh,Chicherin:2019xeg,Hartanto:2019uvl,Abreu:2020cwb,DeLaurentis:2020qle,Chawdhry:2020for}. 
Together with the developments on the IR subtraction side, this has made it possible
to provide the first NNLO studies for production of three photons at the LHC~\cite{Chawdhry:2019bji,Kallweit:2020gcp}.

The breakthrough in the calculation of the relevant massless $2 \to 3$ scattering amplitudes has become possible, on one
hand, due to development of more powerful techniques for the solution of integration-by-parts 
identities (IBPs)~\cite{Tkachov:1981wb,Chetyrkin:1981qh,Laporta:2001dd} which
include new algorithms for their analytical solution~\cite{Gluza:2010ws,Schabinger:2011dz,Ita:2015tya,Larsen:2015ped,Boehm:2017wjc,Boehm:2018fpv,Maierhoefer:2017hyi,Chawdhry:2018awn,Klappert:2019emp,Agarwal:2020dye} and the use of finite fields arithmetic~\cite{vonManteuffel:2014ixa,vonManteuffel:2016xki,Peraro:2016wsq,Peraro:2019svx,Smirnov:2019qkx}, 
and, on the other, due to completion of the calculation of \emph{all} relevant master integrals (MIs) in terms of a well understood
class of functions~\cite{Papadopoulos:2015jft,Gehrmann:2018yef,Chicherin:2018mue,Chicherin:2018old,Chicherin:2020oor}. This has been achieved by the use of 
the method of differential equations~\cite{Bern:1993kr,Kotikov:1990kg,Remiddi:1997ny,Gehrmann:1999as,Papadopoulos:2014lla,Ablinger:2015tua,Primo:2016ebd} augmented by the choice of a canonical basis~\cite{Kotikov:2010gf,Henn:2013pwa}. 
Alternative approaches to the reduction of two-loop amplitudes based on ideas from the
one-loop generalised-unitarity program as well as finite field arithmetic have also been very successful~\cite{Badger:2017jhb,Ita:2015tya,Badger:2018enw,Badger:2019djh,Abreu:2020xvt}.

In this paper, we consider the calculation of two-loop QCD corrections to the production of
a diphoton pair and a jet for the partonic channels $q \bar{q} \to g \gamma \gamma$ and $q g \to q \gamma \gamma$, based on a Feynman diagrammatic approach.
We demonstrate how the use of new algorithms for the reduction of loop integrals and multivariate partial fraction decomposition allows us to compute these two-loop corrections in a rather straightforward way and to produce very
compact and efficient numerical implementations for them.
While it is common practice to compute helicity amplitudes for multi-loop processes, 
the case of diphoton production plus jet does not require us to keep track of the polarisations of the
external photons.
We therefore expect it to be sufficient for near-term phenomenological applications to consider the interference of the two-loop amplitudes with the corresponding tree-level amplitude, summed over colours and
polarisations. We will show that, also in this case, very compact results can be obtained, similar to what can be achieved
for comparable helicity amplitudes. 
Together with the recently computed three-loop QCD corrections to diphoton production~\cite{Caola:2020dfu}, these amplitudes also provide
an essential ingredient towards the calculation of $pp \to \gamma \gamma$ in  N$^3$LO QCD.

The rest of the paper is organised as follows.
In~\refse{se:kinematics} we describe the processes and their kinematics.
In~\refse{se:amplitude} we illustrate the general structure of the
scattering amplitudes, and in particular of the interference terms contributing
to the squared amplitude.
Technical aspects of the diagrammatic calculation, the integral reductions and the multivariate partial decompositions are presented in~\refse{se:technical}.
In~\refse{se:UVrenormIRsubtr} we describe our renormalisation and infrared subtraction
procedures which define the final form of the results.
In~\refse{se:results} we discuss how we optimise our analytic results by using a minimal set of rational functions,
and in~\refse{se:numerical} we present their implementation in a \texttt{C++} numerical code.
We finally draw our conclusions in~\refse{se:conclusions}.

\section{Kinematics and Notation}
\label{se:kinematics}
We consider the production of a pair of photons in association with a gluon in quark-antiquark annihilation
\begin{equation}
\label{eq:qqgaaprocess}
q(p_1) + \bar{q}(p_2) \to g(p_3) + \gamma(p_4) + \gamma(p_5)
\end{equation}
up to two-loop order in massless QCD, \ie corrections up to $\calO(\as^2)$ relative to the tree-level.
We focus here on the scattering process in the physical region, \ie a $2\to 3$ kinematic configuration.
For simplicity, in what follows,
we do not discuss in detail the other partonic subchannel, $qg \to q \gamma \gamma$,
which can be obtained by crossing symmetry according to
\begin{equation}
\label{eq:qgqaaprocess}
q(p_1) + g(p_2)  \to q(p_3) + \gamma(p_4) + \gamma(p_5)\,.
\end{equation}
Analytic results for both channels, with the appropriate identification
of the external momenta as in Eqs.\ \eqref{eq:qqgaaprocess} and \eqref{eq:qgqaaprocess}, 
are provided in the ancillary
files of the arXiv submission of this paper.
Kinematics are fixed by imposing that all  external particles fulfil the on-shell condition
$p^2_i=0$, such that one is left with five independent kinematic invariants, which we choose to be the five adjacent ones,
\begin{gather} 
  s_{12}=(p_1+p_2)^2, \quad  s_{23}=(p_2-p_3)^2, \quad s_{34}=(p_3+p_4)^2, \nonumber \\
  s_{45}=(p_4+p_5)^2, \quad  s_{15}=(p_1-p_5)^2 \,.
  \label{eq:indinvs}
\end{gather}
Note the relative sign between initial- and final-state momenta reflecting our kinematic configuration which implies for the physical region  
$s_{12},s_{34},s_{45} >0$ and $s_{23},s_{15} < 0$.
All the other invariants follow by momentum conservation
and can be derived from the independent ones \refeq{eq:indinvs} via
\begin{align}
  s_{13} &= s_{45}-s_{12}-s_{23},  
& s_{14} &= s_{23}-s_{45}-s_{15}, 
& s_{24} &= s_{15}-s_{23}-s_{34}, \nonumber \\
  s_{25} &= s_{34}-s_{12}-s_{15},  
& s_{35} &= s_{12}-s_{34}-s_{45}\,.
\end{align}
It is also useful to introduce the parity-odd invariant
\begin{equation}
  \label{eq:epsfive}
  \epsfive = 4 \ri \epsilon_{\mu\nu\rho\sigma} p^\mu_1 p^\nu_2 p^\rho_3 p^\sigma_4,
\end{equation}
where $\epsilon_{\mu\nu\rho\sigma}$ is the totally anti-symmetric Levi-Civita symbol.
The invariant $\epsfive$ is related to the determinant of the 
Gram matrix $G_{ij}$ through
\begin{equation}
    (\epsfive)^2 = \Delta \equiv \mathrm{det}G_{ij} =
    \det \left(2 p_i \cdot p_j\right),\, \quad\quad
    i,j \in \left\lbrace1,\ldots,4\right\rbrace,
\end{equation}
which reads explicitly
\begin{equation}
    \Delta =
    (s_{12} s_{23} + s_{23} s_{34} - s_{34} s_{45} + s_{45} s_{15} - s_{15} s_{12})^2 - 
    4 s_{12} s_{23} s_{34} (s_{23} - s_{45} - s_{15})\, .
\end{equation}
As shown in \citere{Byers:1964ryc}, $\Delta<0$ in the physical region,
and $\epsfive = \pm i \sqrt{|\Delta |}$, where the sign depends on the region of phase space.
Since the helicity summed squared matrix elements we consider are even under a parity transformation,
we find it convenient to express our final results in terms of the quantity
\begin{equation}
\tilde{\epsilon}_5 = i \sqrt{ | \Delta | } = i |\epsfive|\,,
\end{equation}
which allows to match the default conventions for the master integrals provided in~\cite{Chicherin:2020oor}.
In the physical scattering region of \refeqs{eq:qqgaaprocess}{eq:qgqaaprocess}
the invariants are constrained to fulfil~\cite{Gehrmann:2018yef}
\begin{equation}
s_{12} \geq s_{34}\,, \quad s_{12} - s_{34} \geq s_{45}\,, \quad 0 \geq s_{23} \geq s_{45}-s_{12}\,, \quad s_{15}^- \leq s_{15} \leq s_{15}^+\,,
\end{equation}
with
\begin{align}
s_{15}^\pm = \frac{1}{(s_{12}-s_{45})^2} &\Big[ s_{12}^2 s_{23} + s_{34} s_{45} (s_{45} - s_{23}) - s_{12} (s_{34} s_{45} + s_{23} s_{34} + s_{23} s_{45})
\nonumber \\
&\pm \sqrt{s_{12} s_{23} s_{34} s_{45} (s_{12} + s_{23} - s_{45})(s_{34} + s_{45} - s_{12})}\, 
\Big]\,.
\end{align}
\section[Structure of the scattering amplitude]{Structure of the scattering amplitude}
\label{se:amplitude}
We start by considering the process in \refeq{eq:qqgaaprocess}.
Its UV-renormalised scattering amplitude can be written as
\begin{equation}
{\rm A}^a_{ij} = \mathbf{T}^a_{ij} \calA^{\mu\nu\rho}
\epsilon_{\mu}^\ast(p_3) \epsilon_{\nu}^\ast(p_4) \epsilon_{\rho}^\ast(p_5) \,
= \mathbf{T}^a_{ij} \calA,
\end{equation}
where we find it convenient to factor out 
the $\mathrm{SU}(3)$ colour generator $\mathbf{T}^a_{ij}$, with
$i$ and $j$ being the colour indices of the quark and anti-quark and
$a$ the colour index of the gluon in the adjoint representation.
$\epsilon^\ast(p_3)$ is the polarisation vector of the outgoing gluon and 
$\epsilon^\ast(p_{4})$, $\epsilon^\ast(p_{5})$ are the ones of the photons.
Clearly, for the $qg$ channel in \refeq{eq:qgqaaprocess}
the same applies upon suitably renaming the external momenta $p_2 \leftrightarrow p_3$ and ommitting the complex conjugate for the polarisation vector of the incoming gluon.
The amplitude $\calA$ stripped of the colour generator $\mathbf{T}^a$
is then perturbatively expanded in the strong coupling constant $\as$ as
\begin{equation}
\label{eq:amplexpansion}
\calA = (4\pi\aew) Q^2_q \sqrt{4\pi\as}
\lb \calA_{0} + \lb\aspi\rb \calA_{1} + \lb\aspi\rb^2 \calA_{2} +
\calO(\as^3)\rb,
\end{equation}
with $\aew$ the fine-structure constant and $Q_q$ the electric charge
of the quarks in units of electron charge.
The decomposition of \refeq{eq:amplexpansion}
fixes the normalisation of the expansion terms $\calA_i$.
The amplitude squared and summed over colours and polarisations
can be expressed as
\begin{equation}
\label{eq:squaredamplitude}
\sum_{ \rm col,pol} | {\rm A}_{ij}^a|^2 = 
\calC
\lb \calW_{00} + \lb\aspi\rb 2\rm{Re}\lsb \calW_{01}\rsb + 
\lb\aspi\rb^2 \lb\calW_{11} + 2\rm{Re}\left[\calW_{02}\right]\rb +
\calO(\as^3)\rb,
\end{equation}
where we introduced the interference terms
\begin{equation}
\label{eq:interferences}
\calW_{ij} = \sum_{\mathrm{pol}} \calA^*_i \calA_j,
\end{equation}
and a global factor $\calC$, that accounts for 
colour sums, which is defined as
\begin{equation}
\label{eq:LOfactor}
\calC= (4\pi)^3 \aew^2 \as (N^2-1) Q^4_q \,. \quad\quad\quad
\end{equation}
Note that all colour and helicity degrees of freedom are summed over in
\refeq{eq:squaredamplitude} rather than performing an average for the initial states.
The perturbative corrections to the expansion coefficients for the amplitudes $\calA_i$, and to their interferences
$\calW_{ij}$, can be decomposed in terms  of their colour and electroweak factors 
which can be expressed as polynomials in  
\begin{equation}
\label{eq:colourfactorsol}
C_A = N, \quad
C_F = \frac{N^2-1}{2N}, \quad
\nfvo = \frac{1}{Q_q} \sum^{n_f}_{i} Q_i, \quad
\nfvt = \frac{1}{Q^2_q}\sum^{n_f}_{i} Q^2_i \,,
\end{equation}
where $C_A$ and $C_F$ are the quadratic Casimirs for the colour group $\mathrm{SU}(N)$,
($N=3$ for QCD), and $n_f$ is the number of (massless) quarks running in closed fermionic loops.
The factors $\nfvo$ and $\nfvt$ correspond to the number of quarks, weighted with their electric charge,
running in closed fermion loops, with one and two external photons attached to, respectively.
The coefficients of various powers of the factors in \refeq{eq:colourfactorsol}
constitute gauge-invariant subsets of the final result, thus it is natural to decompose the
amplitude in terms of them.

\begin{figure}[t]
    \centering
    \includegraphics[scale=0.85]{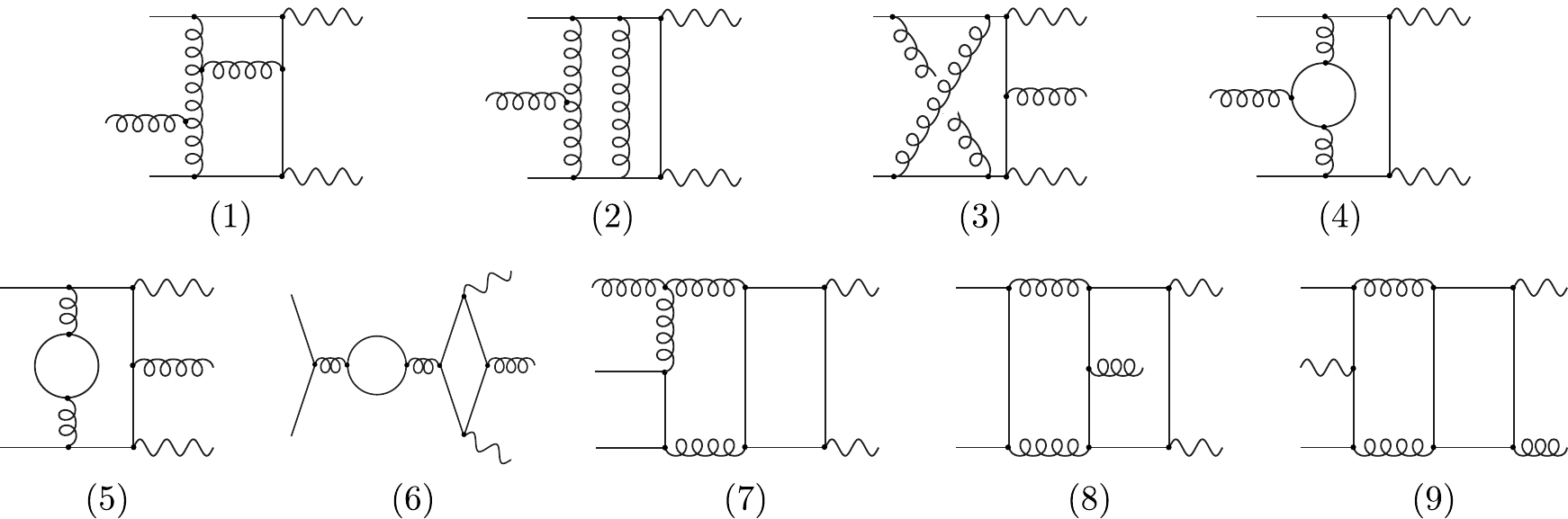}
    \caption{Examples of two-loop diagrams which contribute to the various colour factors in~\refeq{eq:colourfactorstl}. }
    \label{fig:twoloopdiags}
\end{figure}
Within the prescription of \refeqs{eq:interferences}{eq:LOfactor}, the LO result $\calW_{00}$ is a rational function of 
invariants only, while the tree with one-loop interference admits the decomposition
\begin{equation}
\calW_{01} = C_A \calW^{(1)}_{01} + C_F \calW^{(2)}_{01} + \nfvt \calW^{(3)}_{01}
+ n_f \calW^{(4)}_{01}\, .
\end{equation}
The last contribution is introduced by the renormalisation procedure,
see \refse{se:UVrenormIRsubtr} for details.
The tree-two loop interference has a richer structure, which we decide to organise as
\begin{equation}
\calW_{02} = \sum^{10}_{i=1} c_i \calW^{(i)}_{02}\, ,
\end{equation}
with the individual colour factors given by
\begin{align}
c_1 &= C_A^2,   &    c_2 &= C_A C_F,    &  c_3 &= C^2_F,     &&  c_4 = C_A n_f, \nonumber \\  
c_5 &= C_F n_f, &    c_6 &= \nfvt n_f,  &  c_7 &= C_A \nfvt, &&  c_8 = C_F \nfvt,  \nonumber \\ 
c_9 &= \lb 8 C_F - 3 C_A \rb \nfvo,     & c_{10} &= n_f^2, & & &&
\label{eq:colourfactorstl}
\end{align}
where $c_9$ arises from a contraction of type $d_{abc}d_{abc}$ and
$c_{10}$ follows again from ultraviolet (UV) renormalisation.
A selection of representative diagrams contributing to each colour structure
is shown in \reffi{fig:twoloopdiags}, where diagram $(i)$ contributes to the factor
$c_i$ and the colour factor $c_{10}$ follows from diagrams of type $(9)$ as well.
Eventually, one can then expand the polynomial in $N$ for $c_{1,2,3}$ and write
\begin{equation}
C^2_A \calW^{(1)}_{02} + C_A C_F \calW^{(2)}_{02} + \calW^{(3)}_{02} =
N^2 \widetilde{\calW}^{(1)}_{02} + \widetilde\calW^{(2)}_{02} + \frac{1}{N^2} \widetilde\calW^{(3)}_{02},
\end{equation}
where $ \widetilde{\calW}^{(1)}_{02}$ is the leading colour (LC), 
$ \widetilde{\calW}^{(2)}_{02}$ is the next-to-leading colour (NLC) 
and $ \widetilde{\calW}^{(3)}_{02}$ is the next-to-next-to-leading colour (NNLC) contribution.
The expression in terms of Casimir operators can always be recovered through the set of identities
\begin{equation}
N^2 = C^2_A, \quad 1 = C_A^2 - 2 C_A C_F, \quad \frac{1}{N^2} = C_A^2 - 4 C_A C_F + 4 C_F^2\,.
\end{equation}
In this paper we will focus on the calculation of the LC contribution 
$\widetilde{\calW}^{(1)}_{02}$ and the three fermionic contributions 
$ \calW^{(i)}_{02}$, $i=4,5,6$,
which can be obtained by considering only the planar two-loop diagrams.
\section{Diagrammatic calculation and integral reductions}
\label{se:technical}
We perform our calculation in conventional dimensional regularisation (CDR)~\cite{tHooft:1972tcz,Bollini:1972ui}, working in $d = 4-2\eps$ dimensions.
UV and IR singularities will then manifest as poles in the regulator $\eps$.

In order to compute the coefficients $\calW_{00}$, $\calW_{01}$ and $\calW_{02}$, 
we start by producing all
relevant Feynman diagrams with QGRAF~\cite{Nogueira:1991ex}. 
We find $6$ diagrams at tree level, $80$ diagrams at one loop
and $1716$ diagrams at two loops. We use FORM~\cite{Vermaseren:2000nd} to manipulate the diagrams,
interfere them with their tree-level counterparts and perform the relevant Lorentz and Dirac algebra.
Starting at two loops, it is particularly convenient to group the diagrams depending on the graph they can be mapped to,
before performing all symbolic manipulations. This allows us to avoid repeating expensive operations multiple times.
After this step is complete, we can express our two-loop interfered amplitude in terms of scalar Feynman integrals drawn from
two different integral families. We define the integrals as
\begin{equation}
\mathcal{I}^{\rm fam}_{n_1,...,n_{11}} = 
e^{2 \epsilon \gamma_E }\, \int \prod_{i=1}^2 \left(\frac{d^d k_i}{i \pi^{d/2}}\right) \frac{1}{D_1^{n_1} ... D_{11}^{n_{11}} }\,,
\end{equation}
where $\gamma_E \sim 0.5772$ is the Euler constant and,
following the notation of~\cite{Gehrmann:2018yef,Chicherin:2018mue}, we define the two families in \refta{tab:topos}.

\begin{table}[t]
\begin{center}
\begin{tabular}{| c | l | l|}
\hline
Prop.\ den.\ & Family A & Family B  \\
\hline
\hline
$D_1$ & $ k_1^2$ &                                        $k_1^2$   \\
$D_2$ & $(k_1+p_1)^2$ &                              $(k_1-p_1)^2$  \\
$D_3$ & $(k_1+p_1+p_2)^2$ &                      $(k_1-p_1-p_2)^2$   \\
$D_4$ & $(k_1+p_1+p_2+p_3)^2$ &              $(k_1-p_1-p_2-p_3)^2$   \\
$D_5$ & $k_2^2 $ &                                        $k_2^2 $   \\
$D_6$ & $(k_2+p_1+p_2+p_3)^2$ &               $(k_2-p_1-p_2-p_3-p_4)^2$   \\
$D_7$ & $(k_2+p_1+p_2+p_3+p_4)^2$ &       $(k_1-k_2)^2$   \\
$D_8$ & $(k_1 - k_2)^2$ &                               $(k_1-k_2+p_4)^2$   \\
$D_9$ & $(k_1+p_1+p_2+p_3+p_4)^2$ &       $(k_2-p_1)^2$   \\
$D_{10}$ & $(k_2 + p_1 )^2$ &                        $(k_2-p_1-p_2)^2$   \\
$D_{11}$ & $(k_2 +p_1 +p_2)^2$ &                 $(k_2-p_1-p_2-p_3)^2$   \\
\hline 
\end{tabular}
\caption{Definition of the two integral families used in the calculation. We note that also both the non-planar hexagon-box and the double-pentagon topologies can be described by family $B$ and crossings thereof.} \label{tab:topos}
\end{center}
\end{table}

As a first step to simplify our interference terms, we use Reduze\;2~\cite{vonManteuffel:2012np,Studerus:2009ye} 
to search for symmetry relations among the different scalar integrals in the two topologies
in Table~\ref{tab:topos} and their crossings. This allows us to substantially reduce the number of different integrals 
that we need to reduce to master integrals.
First, we collect the integrals that are required to express the coefficients of  \emph{any} of the colour factors \eqref{eq:colourfactorstl}.
We find that after applying the symmetrisations above, and modulo crossings of the external legs, the
interference of the two-loop amplitude with the tree-level amplitude requires the reduction of $1811$ integrals in family $A$
and $508$ integrals in family $B$, which include all the integrals required to compute 
both the hexagon-box and the double-pentagon topologies.
In both families we need at most rank-5 integrals (up to 5 irreducible scalar products in the numerator).
From this point on, we focus only on the planar colour factors and therefore only on integrals belonging to family $A$.
We performed their reduction using both Kira~\cite{Maierhoefer:2017hyi,Maierhofer:2018gpa,Klappert:2020nbg}
and an in-house implementation, Finred, which employs finite field
sampling, syzygy techniques, and denominator guessing, see \cite{Agarwal:2020dye,Heller:2021qkz} for more details.
It is worth stressing that the planar reductions up to rank 5 did not constitute an issue for either program here, and took e.g.\ $40$ hours on $36$ cores with Kira and a similar runtime with Finred.
The reduction lists for family $A$
produced in this way are not extremely complicated, with a size of around $390$ MB in total. 
Note that we performed the reduction directly in terms of the pre-canonical set of master integrals defined in~\cite{Gehrmann:2018yef}.
We stress here that these lists do not include the crossings necessary to reduce all diagrams.

To arrive at a complete set of reduction identities and to render their inclusion as simple as possible, we proceed as follows.
The integration-by-parts solvers deliver each integral coefficient as a rational function in a common-denominator representation.
We find it useful to convert the rational functions to a partial fraction decomposed form.
Due to the choice of master integrals, we encounter only irreducible denominator factors which depend on either $d$ or the kinematic variables.
For the $d$ dependence of the rational functions, a simple univariate partial fraction decomposition is sufficient.
In contrast, the decomposition involving the kinematic denominators is computationally non-trivial.
We encounter the irreducible denominator factors
\begin{align}
\label{eq:denoms}
\{d_1,\ldots,d_{25}\} &= \{ 
    s_{12},
    s_{23},
    s_{34},
    s_{45},
    s_{51},
    s_{12} + s_{23} - s_{45},
    s_{23} - s_{45} - s_{51},
    s_{23} + s_{34} - s_{51}, \nonumber \\
&\quad\;\;\;   s_{12} - s_{34} + s_{51},
    s_{12} - s_{34} - s_{45},
    s_{12} + s_{23},
    s_{12} - s_{34},
    s_{23} + s_{34},
    s_{12} - s_{45}, \nonumber \\
&\quad\;\;\;   s_{23} - s_{45},
    s_{12} + s_{23} - s_{34} - s_{45},
    s_{34} + s_{45},
    s_{23} - s_{51},
    s_{34} - s_{51}, \nonumber \\
&\quad\;\;\;   s_{12} + s_{23} - s_{45} - s_{51},
    s_{23} + s_{34} - s_{45} - s_{51},
    s_{12} + s_{51},
    \nonumber \\
&\quad\;\;\;  s_{12} - s_{23} - s_{34} + s_{51}, 
    s_{12} - s_{34} - s_{45} - s_{51},
    s_{45} + s_{51}
\}.
\end{align}
For them, we employ the \texttt{MultivariateApart} package \cite{Heller:2021qkz} to perform a multivariate partial fraction decomposition; see also 
\cite{Pak:2011xt,Abreu:2019odu,Boehm:2020ijp} for related decomposition techniques.
The decomposition of \cite{Heller:2021qkz}  is based on replacing all irreducible denominator factors
$d_k(\{s_{ij}\})$ in a given expression according to
\begin{align}
\label{eq:qidef}
    \frac{1}{d_k(\{s_{ij}\})} = q_k,
\end{align}
and reducing the resulting polynomial with respect to the ideal generated by
\begin{align}
\label{eq:idealgens}
    \{ d_1(\{s_{ij}\}) q_1 - 1, \ldots, d_{25}(\{s_{ij}\}) \}
\end{align}
using a suitable monomial ordering.
Due to the specific structure of our denominator list, the monomial block ordering of \cite{Heller:2021qkz} coincides with a lexicographic ordering of the $q_i$ before any $s_{ij}$ are considered, which produces a Le{\u\i}nartas decomposition~\cite{Leinartas:1978pf,Raichev:2012pf}.
Note that we included only the local denominators of each coefficient in the initial partial fraction decomposition.
This run took a couple of days and reduced the size of the reduction list by almost a factor $5$, down to around $80$ MB.
In particular, the reduction in complexity is particularly pronounced for the most complicated rank-5 identities, for which up to a factor of $40$ reduction in size is seen.
While it has been known for a long time that integration-by-parts identities can become substantially simpler even using naive variants of multivariate partial fraction decompositions, we would like to point out a systematic study of the impact of these new algorithms on the size of the reduction identities which has recently appeared in~\cite{Boehm:2020ijp}.
Starting from these substantially simpler identities, we perform the relevant crossings and then a second partial fraction decomposition.
The second partial fraction decomposition employs a \emph{global} Gr\"obner basis for the ideal \refeq{eq:idealgens}, taking into account the denominator factors of all coefficients at once.
We emphasize that this method allows to decompose terms of a sum locally, but still ensures a globally unique representation of the rational functions in the results.
With this multi-step procedure, we obtained all identities and their crossings in a form suitable for insertion in the diagrammatic calculation, while keeping the dimension of the expressions under control at each step.
After insertion of the reduction identities into the interference terms, a final quick last partial fraction decomposition handles the remaining few additional denominator factors.
In practice, we find the partial fraction decomposition of the original, uncrossed identities to be the most time-consuming step, which nevertheless could be handled in a couple of days. 
The remaining steps to arrive at the reduced and fully partial fractioned interference terms took a few hours.

In order to optimize our results for numerical stability, we found it useful to tune our monomial ordering for the partial fraction decomposition in such a way that  spurious poles and unnecessarily high powers of the denominators are avoided.
This choice of ordering was performed for each colour factor and loop order separately.
In this way, we obtained a rather compact expression for the two-loop interference terms in terms of canonical master integrals, with full dependence on the space-time dimensions $d$.
As we will see in \refse{se:results}, by substituting the explicit results for the master integrals, expanding about $d=4$, subtracting the poles, and properly simplifying the remaining rational functions, we are able to obtain extremely compact, fully analytic results for the finite remainders of the interference terms.

\section{Renormalisation and infrared factorisation}
\label{se:UVrenormIRsubtr}
We work in fully massless QCD and keep the number of light fermion flavours $n_f$ generic; we do not consider any loop corrections due to massive quarks.
We perform UV renormalisation in the standard $\overline{\mathrm{MS}}$ scheme,
which allows one to express renormalised amplitudes in terms of 
unrenormalised ones by simply replacing the bare coupling constant $\alpha^b_s$
with the running coupling $\alpha_s(\mu^2)$, evaluated at the scale $\mu^2$,
\begin{equation}
\alpha_s^b \mu^{2\eps}_0 S_\eps = 
\as \mu^{2\eps}\left[1 - \frac{\beta_0}{\eps}\lb\asontwopi\rb + 
\lb\frac{\beta_0^2}{\eps^2} - \frac{\beta_1}{2\eps}\rb \lb\asontwopi\rb^2 +
\calO(\as^3)\right],
\end{equation}
where $S_\eps = (4\pi)^\eps e^{-\eps\gamma_E}$, and $\beta_0$ and $\beta_1$ denote the first two perturbative orders of the QCD beta function,
\begin{equation}
    \beta_0 = \frac{11}{6}C_A - \frac{2}{3}T_r n_f, \quad
    \beta_1 = \frac{17}{6}C^2_A - \frac{5}{3}C_A T_r n_f - C_F T_r n_f,
\end{equation}
with $T_r = 1/2$. 
The renormalised interference terms in \refeq{eq:squaredamplitude} are explicitly obtained as
\begin{align}
\calW_{01} &= S^{-1}_\eps\calW^b_{01} - \frac{\beta_0}{2\eps} \calW_{00}, \nonumber \\
\calW_{02} &= S^{-2}_\eps\calW^b_{02} - \frac{3\beta_0}{2\eps} S^{-1}_\eps\calW^b_{01}
- \lb\frac{\beta_1}{4\eps} - \frac{3\beta_0^2}{8\eps^2}\rb \calW_{00}, 
\label{eq:renorminterf}
\end{align}
where with a superscript $b$ we denote bare quantities.
After UV renormalisation the results in \refeq{eq:renorminterf} contain only IR poles.
We subtract them according to Catani's factorisation formula~\cite{Catani:1998bh}, which 
makes it possible to reorganise the interference terms as
\begin{align}
\calW_{01} &= I_1(\eps,\mu^2) \calW_{00} + \calW^\fin_{01}, \nonumber \\
\calW_{02} &= I_1(\eps,\mu^2) \calW_{01} + I_2(\eps,\mu^2) \calW_{00} + \calW^\fin_{02},
\label{eq:CataniSubtraction}
\end{align}
where the action of the operators $I_1$ and $I_2$ produce the complete IR structure of the
renormalised amplitudes.
The finite remainders 
$\calW^\fin_{01}$, $\calW^\fin_{02}$ will be the main result of our paper.
Before discussing these in detail, let us show the explicit formulae 
for the Catani operators for the process under consideration.\footnote{
Note that the form of $I_1$ and $I_2$ is dictated entirely by the presence of a $q\bar{q}$ pair and one gluon as coloured external states.} By direct calculation it is easy to see that
for the $q\bar{q}$ channel
\begin{align}
    I_1(\eps,\mu^2) = \frac{\mathrm{e}^{\eps\gamma_{E}}}{2\Gamma(1-\eps)}
    \bigg[ &\lb C_A - 2 C_F\rb \lb \frac{1}{\eps^2} + \frac{3}{2\eps} \rb
    \lb-\frac{\mu^2}{s_{12}}\rb^\eps \nonumber \\
    -& \lb C_A \lb \frac{1}{\eps^2} + \frac{3}{4\eps} \rb + \frac{\gamma_g}{2\eps} \rb
    \lb \lb-\frac{\mu^2}{s_{13}}\rb^\eps + \lb-\frac{\mu^2}{s_{23}}\rb^\eps \rb\bigg],
\label{eq:CataniIone}
\end{align}
where $\gamma_g = \beta_0$, and
\begin{align}
    I_2(\eps,\mu^2) = & -\frac{1}{2} I_1(\eps,\mu^2)\lb I_1(\eps,\mu^2) + 2\frac{\beta_0}{\eps} \rb 
    \nonumber \\ 
    +&
    \mathrm{e}^{-\eps\gamma_{E}}
    \frac{\Gamma(1-2\eps)}{\Gamma(1-\eps)}\lb\frac{\beta_0}{\eps}+K\rb I_1(2\eps,\mu^2) +
    \frac{\mathrm{e}^{-\eps\gamma_{E}}}{4\eps\Gamma(1-\eps)} H_2,
\end{align}
where $K$ is universal,
\begin{equation}
    K = \lb \frac{67}{18} - \frac{\pi^2}{6}\rb C_A - \frac{10}{9} T_r n_f,
\end{equation}
and $H_2$ is a renormalisation and process-dependent factor, in our case given by
\begin{equation}
H_2 = 2 H_{2,q} + H_{2,g},
\end{equation}
with $H_{2,q}$ and $H_{2,g}$ which read explicitly~\cite{Garland:2002ak}
\begin{align}
    H_{2,g} =& \lb \frac{\zeta_3}{2} + \frac{5}{12} + \frac{11\pi^2}{144} \rb C^2_A +
    \frac{20}{27}T_r^2 n_f^2 - \lb \frac{\pi^2}{36} + \frac{58}{27}\rb C_A T_r n_f + C_F T_r n_f, \\
    H_{2,q} =& \lb -\frac{3}{8} + \frac{\pi^2}{2} - 6 \zeta_3 \rb C^2_F + 
    \lb \frac{245}{216} - \frac{23\pi^2}{48} + \frac{13\zeta_3}{2} \rb C_A C_F \nonumber \\ 
    +& \lb\frac{\pi^2}{12} - \frac{25}{54}\rb C_F T_r n_f\, .
\end{align}
In order to obtain the corresponding expressions for the $qg$ channel,
it is enough to swap the indices $2\leftrightarrow 3$ in \refeq{eq:CataniIone} 
and the rest follows.

Since we are working in CDR, we carry out the computation of the LO term $\calW_{00}$
retaining exact $d$ dependence and we expand $\calW_{01}$ up to $\calO(\eps^2)$.
This is necessary for the correct subtraction of the IR poles according to
\refeq{eq:CataniSubtraction}.

After UV renormalisation and IR subtraction the one-loop finite remainder takes the form:
\begin{equation}
\label{eq:oneloopfinite}
    \calW^\fin_{01} = 
    C_A \calW^{(1),\fin}_{01} +
    C_F \calW^{(2),\fin}_{01} +
    \nfvt \calW^{(3),\fin}_{01} + 
    n_f \calW^{(4),\fin}_{01},
\end{equation}
where, as anticipated in \refse{se:amplitude}, the last term on the right-hand side
originates from the renormalisation and subtraction procedures and 
it is proportional to the LO result.
As for the two-loop finite remainder, we find it convenient to cast it in the form
\begin{equation}
\label{eq:twoloopfinite}
    \calW^\fin_{02} = 
    \sum_{i=1}^3 \tilde{c}_i \widetilde{\calW}^{(i),\fin}_{02} + 
    \sum_{i=4}^{10} c_i \calW^{(i),\fin}_{02},
\end{equation}
where the colour factors $c_{4,\ldots,10}$ are the same as defined in
\refeq{eq:colourfactorstl} and we introduced
\begin{equation}
    \tilde{c}_1 = N^2, \quad
    \tilde{c}_2 = 1, \quad
    \tilde{c}_3 = N^{-2}\, .
\end{equation}
The factors $\tilde{c}_{1,2,3}$ clearly organise the colour tower, 
while $c_{10}$ is introduced by UV renormalisation and IR subtraction.
As we have already argued at the end of Sec.~\ref{se:amplitude}, the coefficients 
$\widetilde{\calW}^{(1),\fin}_{02}$, $\calW^{(4),\fin}_{02}$, $\calW^{(5),\fin}_{02}$ and 
$\calW^{(6),\fin}_{02}$ entail only planar two-loop corrections.
These four finite contributions constitute the main object of this paper.

A first test that we performed on our results at the analytical level,
is that we reproduce all the $\eps$ poles of the renormalised interference terms, 
at one and two loops, as predicted by 
Catani's factorisation formula in \refeq{eq:CataniSubtraction}.
\section{Exploiting linear dependencies of rational functions}
\label{se:results}
The LO contribution defined in \refeq{eq:squaredamplitude} for process
\refeq{eq:qqgaaprocess} has a very
compact expression, which, computed in CDR, reads
\begin{align}
\calW_{00} =& \frac{8s_{12}}{s_{14}s_{15}s_{24}s_{25}} \big(
s^2_{13}+s^2_{23}
+ \eps(2-\eps)(s_{14}^2+s_{15}^2+s_{24}^2+s_{25}^2) \nonumber \\
+& 2\eps(1-\eps)s_{13} s_{23} + \eps(3+\eps)(s_{14}-s_{24})(s_{15}-s_{25})\big) 
\nonumber \\
+& \big( 3\to 4, 4 \to 5, 5 \to 3 \big)
+  \big( 3\to 5, 5 \to 4, 4 \to 3 \big), \label{eq:treelevel}
\end{align}
where we kept full $\eps$-dependence and in the last row we denote 
cyclic permutations of the indices labelling the kinematic invariants.
An equivalent expression holds for process~\refeq{eq:qgqaaprocess} 
upon replacing $2\leftrightarrow 3$ in the indices labeling the invariants.

Starting already from one-loop, the explicit expression of the interference
terms become too lengthy to be shown here. 
Therefore, we limit ourselves to describing the functional form of our results
and the approach we adopted to obtain particularly compact expressions.

Both at one and two loops, the finite remainders will be a linear combination
of transcendental functions multiplied by rational functions.
The physical one-loop finite remainders, \ie the coefficients of $\calO(\eps^0)$,
contain only simple logarithms and dilogarithms and 
they are free of the parity odd-invariant $\epsfive$.
At the two-loop order, as well as for the higher $\eps$ powers of the one loop result,
the class of transcendental functions needs to be extended
beyond classical polylogarithms and $\epsfive$ enters explicitly.
We employ the representation in terms of so-called Pentagon Functions presented in~\cite{Chicherin:2020oor}\footnote{For further details about the definition and implementation of the pentagon functions see \cite{Chicherin:2020oor}.}.
Thus we write a generic finite remainder as
\begin{equation}
\label{eq:ratpentafunscombo}
    \calW^{(m),\fin}_{0n} = 
    \sum_{k} r_{k}(\left\lbrace s_{ij},\epsfive\right\rbrace) f_k(\left\lbrace s_{ij},\epsfive\right\rbrace),
\end{equation}
where $m$ denotes some given colour factor, $n=1,2$, $r_k$ is a rational function
of the kinematic invariants and $f_k$ indicates a pentagon function.
Equation \refeq{eq:ratpentafunscombo} holds formally at LO, $n=0$, as well, upon
putting $f_k = 1$. We remind the reader that at this step 
we represent each rational function $r_k$ in its partial fractions decomposed form.
We further stress that the algorithm presented in~\cite{Heller:2021qkz} avoids the presence
of spurious denominators, therefore our final expressions contain all and only the physical
denominators of the amplitudes.

Despite that the partial fraction decomposition allows us to obtain rather compact results already at this level, 
we find that the representation in \refeq{eq:ratpentafunscombo} is not the most compact representation yet.
This is due to the fact that the rational functions $r_k$ are not
linearly independent and, in fact, the set of independent rational functions is substantially smaller.
The minimal set of rational functions can be obtained as follows.
First of all, in order to guarantee a globally unique representation of all rational functions, we employ a partial fraction decomposition with respect to a common Gr{\"o}bner basis.
As a result, each rational function $r_k$ is represented as a polynomial of maximum degree $p$ in the 5 kinematic variables $s_{ij}$ \eqref{eq:indinvs} and the 25 inverse denominators $q_i$ \eqref{eq:qidef},
\begin{align}
\label{eq:rkpartfrac}
    r_k &= \sum_{m_1 + \ldots +m_n \leq p} a_{k,m_1\ldots m_{30}} M_{m_1\ldots m_{30}}, \text{~where~} M_{m_1\ldots m_{30}} \equiv q_1^{m_1}\cdots q_{25}^{m_{25}} s_{12}^{m_{26}}\cdots s_{51}^{m_{30}}
\end{align}
and the coefficients $a_{k,m_1\ldots m_{30}}$ are rational numbers.
We emphasize that only certain $M_{k,m_1\ldots m_{30}}$ occur as irreducible monomials in our partial fractioned results; in practice it is convenient to enumerate only those.
We are interested in determining linear relations between different rational functions, \begin{equation}
\label{eq:rklindep}
    0 = \sum_k b_k r_k\,.
\end{equation}
and employing them to re-express all $r_k$ in terms of a linearly independent subset of them.
By inserting the partial fractioned form \eqref{eq:rkpartfrac} and observing that the monomials $M_{m_1\ldots m_{30}}$ are independent, we obtain a system of linear relations for the $b_k$,
\begin{equation}
\label{eq:bsystem}
    0 = \sum_k a_{k,m_1\ldots m_{30}} b_k
\end{equation}
for each of the irreducible monomials $M_{m_1\ldots m_{30}}$.
In our case, there are many more monomials than rational functions, 
and the system is over-constrained.
Nevertheless, many of the equations turn out to be linearly dependent, 
allowing us to find a solution to the system in terms of a basis of 
independent rational functions.
Note that this can be achieved by a row reduction of a matrix of rational numbers, 
which can easily be done e.g.\ with a finite field solver like \texttt{Finred}.
Similarly to what happens for the reduction to master integrals, 
there is not a unique solution for the basis of rational functions 
and a more natural choice can lead to more compact results for the 
scattering amplitude.
We find that very compact analytic expressions can be found 
by simplifying the rational functions for each loop order and colour factor
separately. In practice, we order the rational functions according to the number
of monomials they contain, and at each step we remove the one  
with the largest number.

This approach makes it possible to drastically reduce the size of the final results. 
As an example, for the LC coefficient $\widetilde{\calW}^{(1),\fin}_{02}$ 
we start with an expression of around 64 MB 
which, after moving to a basis of independent rational functions, can be
reduced to around 3 MB.
This makes it possible not only to have a more compact result,
but, most importantly, a more efficient and potentially more stable numerical implementation. 
Clearly, we cannot demonstrate that a more compact representation does not exist for a different choice of independent rational functions. On the contrary, we expect that
more compact expressions could be obtained if one does not insist on using
a set of independent kinematics invariants, see for example Eq.~\refeq{eq:treelevel} and Ref.~\cite{DeLaurentis:2020qle}.
Nevertheless, our current representation is more than suitable for practical use.

As last a remark, we stress that the procedure above does not produce the minimal number of 
rational functions. Indeed, starting from the results
simplified according to the procedure above, we can attempt to further
relate rational functions
across different colour structures and different loop orders. In this way,
more relations can be found and a minimal set of rational functions
can be identified. We find this second representation particularly
suitable for implementation of the interference terms in a numerical code,
as described in the following section.
We provide analytical expressions for various colour factors at the different
loop orders with the arXiv submission of this manuscript.

\section{Numerical implementation}
\label{se:NumericalImplementation}
We implemented our results in the \texttt{C++} numerical code
\texttt{aajamp}, which we distribute
through the git repository~\cite{aajamplib}; it can be easily downloaded with
\begin{equation*}
    \texttt{git clone https://gitlab.msu.edu/vmante/aajamp.git}
\end{equation*}
Details on the installation procedure and usage of the package are given in
the git repository.
The main purpose of the code is to evaluate the finite remainders $\calW_{00}$, 
all $\calW^{(i),\fin}_{01}$ of \refeq{eq:oneloopfinite} and 
$\calW^{(i),\fin}_{02}$ of \refeq{eq:twoloopfinite} for 
$i\in\left\lbrace1,4,5,6,10\right\rbrace$.
The evaluation of all transcendental functions in our code, both at one and two-loop
level, is handled by the \texttt{PentagonFunctions-cpp} library of
\cite{Chicherin:2020oor,PentagonFunctions:cpp,Li2pp}.
In order to implement our formulae in an efficient fashion we adopted the
optimised code generation provided by \texttt{FORM}~\cite{Kuipers:2013pba}.

We then checked the implementation of our LO and NLO results against 
\texttt{OpenLoops\;2} \cite{Buccioni:2019sur} and find agreement.
\begin{table}[t]
\begin{center}
\begin{tabular}{| c | c | c |}
\hline
$\calW^{(i)}_{0n}$     & $q\bar{q}\to g \gamma\gamma$  & $q g\to q \gamma\gamma$  \\
\hline \hline
$\calW_{00}$   & $5.31483213538515820$ & $5.70333370484282831$\\
\hline
$\calW^{(1),\fin}_{01}$   & $ 20.6589111571951634$   & $-4.61720250742111915$\\
$\calW^{(2),\fin}_{01}$   & $-22.5582575628444708$   & $-12.1699155553812286$\\
$\calW^{(3),\fin}_{01}$   & $ 0.990965411280490205$  & $-1.87586195456244775$\\
$\calW^{(4),\fin}_{01}$   & $-0.654985891275428056$  & $-0.186734013704948326$\\
\hline
$\calW^{(1),\fin}_{02}$   & $ 109.027607189124197$   & $-12.3348917323658682$\\
$\calW^{(4),\fin}_{02}$   & $-29.2470156809751600$   & $7.70946078306355620$\\
$\calW^{(5),\fin}_{02}$   & $ 14.1633315227102639$   & $6.45363246183533512$\\
$\calW^{(6),\fin}_{02}$   & $-0.52366019039559053$   & $1.79398590828584226$\\
$\calW^{(10),\fin}_{02}$  & $0.607997053766843543$   & $-0.66761299936404394$ \\
\hline
\end{tabular}
\caption{Benchmark results for the tree-level, one-loop and two-loop interference terms
for the partonic channels $q\bar{q}$ and $qg$ for the kinematic point in \refeq{eq:benchpoint}.
Only the real part of $\calW^{(i)}_{0n}$ is presented.} \label{tab:benchmark}
\end{center}
\end{table}

In \refta{tab:benchmark} we provide benchmark results for a kinematic point 
in the physical region defined by
\label{se:numerical}
\begin{equation}
\label{eq:benchpoint}
    s_{12} = 157, \quad
    s_{23} = -43, \quad
    s_{34} = 83,  \quad
    s_{45} = 61,  \quad
    s_{15} = -37, \quad
    \mu^2 = 100\, .
\end{equation}

To assess the performance of our code, we measured the evaluation time
of the NLO and NNLO results in double precision on a single
Intel i7-9750H CPU @ 2.60GHz core using \texttt{gcc} 9.3.0 for a distribution of physical points in phase space, which we describe in more detail below.
We find an average evalution time of $5.2\times 10^{-2}$ ms and 1.2 s per phase space point
for the NLO and NNLO contributions, respectively.
We note that most of the evaluation time for a given point is spent 
on the computation of the pentagon functions.

The \texttt{aajamp} package allows for numerical evaluations in double and
quadruple floating-point precision, \ie 16 and 32 decimal digits representation.
The quadruple precision evaluation is adapted to follow the strategy of the 
\texttt{PentagonFunctions-cpp} library, 
which utilizes the \texttt{qd} library \cite{Hida00quad-doublearithmetic:}
for higher-precision arithmetic.\footnote{Strictly speaking the 
32 decimal digits representation of \texttt{qd} is based on a double-double
arithmetic.}
Therefore, \texttt{aajamp} relies on \texttt{qd} as well.
We note that the \texttt{PentagonFunctions-cpp} and \texttt{qd} libraries 
offer octuple precision, but we reckon that quadruple precision is sufficient for
phenomenological applications.

\begin{figure}[t]
    \centering
    \includegraphics[scale=0.75]{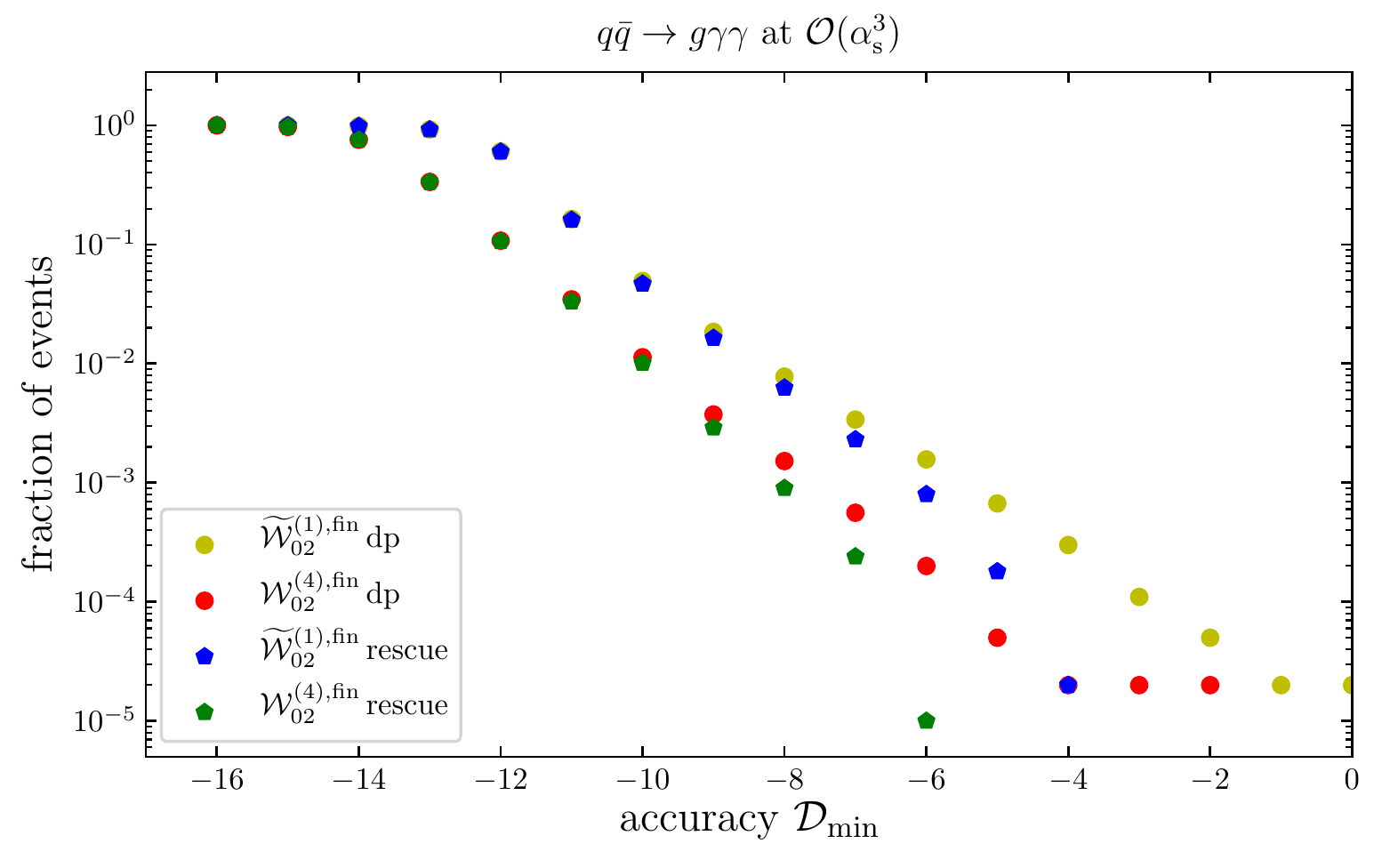}
    \caption{Probability of finding an event with an instability level 
    $\mathcal{D}$ defined in \refeq{eq:instabilitydefinition} larger than $\mathcal{D}_{\mathrm{min}}$
    for the two colour factors $\calW^{(1),\fin}_{02}$ 
    and $\calW^{(4),\fin}_{02}$. The labels ``dp" and ``rescue" are described in the text.
    Events were generated according to \refeq{eq:events}.}
    \label{fig:accuracy}
\end{figure}

Although the user can choose at the interface level between a purely double or 
quadruple arithmetic evaluation, we have implemented a primitive precision
control system.
This is motivated by the fact that individual terms appearing in the squared matrix elements can develop spurious singularities inside the physical phase space, 
leading to more or less severe numerical instabilities.
These are typically associated with small Gram determinants,
or in our case, with denominators in the rational functions which become small
compared to the typical energy scale of the process.
We observe that in our final expressions, the parity odd invariant $\epsfive$ 
does not appear in any denominator and, correspondingly, small Gram determinants are not of prior concern.
Instead, we focus on small denominators in the rational functions, which are not associated with a physical singularity, but lead to a major source of numerical instabilities in the evaluations.
Let us stress that, in principle, also other types of numerical instabilities could arise, but our denominator based analysis indeed works well in practice, as we will show.
Examples of these spurious singularities arise in events with all particle
having a large transverse momentum and entailing
collinear photon pairs or collinear gluon-photon pairs.
We therefore find it natural to activate a quadruple precision evaluation if
any of the 25 denominator factors becomes smaller than a given threshold $\chi$, \ie if
\begin{equation}
\label{eq:stabthresh}
    d_{i} < \chi s_{12}, \quad 
    i \, \in \lbrace 1,\ldots,25 \rbrace\, .
\end{equation}
The threshold $\chi$ can be tuned by the user as detailed in one of the examples
in our git repository.
As can be seen in \reffi{fig:accuracy}, this precision-control system significantly improves the reliability of the numerical evaluations.
Here, we assess the level of numerical stability via a rescaling test, \ie 
we consider the quantity $\mathcal{D}$ defined as
\begin{equation}
\label{eq:instabilitydefinition}
    \mathcal{D} = \log_{10}
    \left|1-\xi^n \frac{\calW^{(i)}_{02}(\xi s_{ij},\xi \mu^2)}{\calW^{(i)}_{02}(s_{ij},\mu^2)} \right|,
\end{equation}
where we perform two evaluations of the colour factor $\calW^{(i),\fin}_{02}$ with input kinematics
$(s_{ij},\mu^2)$ and with the same kinematics rescaled by a factor $\xi$. 
The extra factor $\xi^n$ accounts for the mass dimensionality of $\calW^{(i),\fin}_{lm}$, 
and in our case we just have $n=1$.
The definition of $\mathcal{D}$ in \refeq{eq:instabilitydefinition}
intuitively provides an indicator for the number of digits one 
can trust for a given computation, thus we identify it as our level of instability.
In \reffi{fig:accuracy} we show a cumulative histogram of 
phase-space points, which provide an evaluation of $\calW^{(i),\fin}_{02}$ with an
instability $\mathcal{D}$ larger than $\mathcal{D}_{\mathrm{min}}$.
We plot two selected colour factors, both evaluated in pure double precision, 
labelled as ``dp'', and with the precision control system activated, labelled as ``rescue''
where the threshold $\chi$ has been set to $10^{-4}$. We stress that only those points
which fulfil the condition in \refeq{eq:stabthresh} are evaluated with higher precision.
We generated $10^5$ uniformly distributed events with \texttt{Rambo}~\cite{KLEISS1986359}
subject to the constraints
\begin{equation}
\label{eq:events}
    E_{\mathrm{com}} = 1 \, \mathrm{TeV}, \quad
    p_{\mathrm{T},g} > 30 \, \mathrm{GeV}, \quad
    p_\mathrm{T,\gamma_1} > 30\, \mathrm{GeV}, \quad
    p_\mathrm{T,\gamma_2} > 30\, \mathrm{GeV},
\end{equation}
where $E_{\mathrm{com}}$ is the energy in rest frame of the colliding partons and 
$p_{\mathrm{T},i}$ is the transverse momentum of particle $i$.
One can see that for this ensemble of events, the precision control system is able to
capture and cure the worst instabilities, effectively guaranteeing a very good 
level of accuracy.
Once again, we stress that a more sophisticated stability system could be
devised in principle, especially in regions of soft or collinear emissions. 

\section{Conclusions}
\label{se:conclusions}
In this paper, we  presented the calculation of the leading colour and light fermionic 
two-loop corrections
for the production of two photons and a jet in quark-antiquark annihilation and 
quark-gluon scattering.
This calculation has been made possible by the combination of state-of-the-art
techniques for the reduction of multiloop Feynman integrals, new ideas
about their representations in terms of multivariate partial fractions,
and recent results for the relevant master integrals.
In particular, we have shown how the spin summed interference between the two-loop and the tree amplitude can be computed from its Feynman diagrammatic
representation, resulting in very compact analytic expressions.
We have checked that our two-loop corrections display the
correct pole structure, as first predicted by Catani for QCD amplitudes
at this perturbative order. 

In order to demonstrate the flexibility and usability of our result,
we have also implemented the tree-level, one-loop and two-loop finite remainders 
in a \texttt{C++} library, providing all $\epsilon$ expansions through to transcendental weight four. 
Our library links against the \texttt{PentagonFunctions-cpp} library and allows the user to evaluate the loop corrections in double and quadruple precision.
We have also introduced a simple precision control system that allows the code to identify phase space points prone to loss of precision, such that quadruple precision evaluations are restricted to a minimum.
We envisage that the algorithms developed for this calculation can be
extended to solve future cutting-edge problems in
the computation of multiloop multileg scattering amplitudes
relevant for collider physics phenomenology.

\acknowledgments
We thank Fabrizio Caola and Tiziano Peraro for various discussions on different
aspects of the calculation. 
We further thank Fabrizio Caola for comments on the manuscript.
FB is thankful to Jean-Nicolas Lang for useful suggestions 
on several aspects of the numerical code implementation.
We thank Alexander Mitov and Rene Poncelet for private communications and for sharing
numerical results on their calculation prior to publication.
We thank Vasily Sotnikov for clarifications on the results published in \cite{Chicherin:2020oor}.
The research of FB is supported by the ERC Starting Grant 804394 HipQCD.
BA and AvM are supported in part by the National Science Foundation through Grant 2013859.
LT is supported by the Royal Society through Grant URF/R1/191125.

\bibliographystyle{JHEP}
\bibliography{qqaag_lc}

\end{document}